\documentclass[journal]{IEEEtran}
\IEEEoverridecommandlockouts

\usepackage{cite}
\usepackage{graphicx}
\usepackage{amsmath,amssymb,amsfonts}
\usepackage{algorithmic}
\usepackage{textcomp}
\usepackage{xcolor}
\usepackage{lineno}

\graphicspath{{figures/}{../figures/}}

\begin{document}

\title{Heterogeneous-Gradient Phase--Polarization Alignment and Maximal-Ratio Weight Allocation for Multi-Aperture Coherent FSO Reception}

\author{Cheng~Chen,~Tong~Luo,~Jiayin~Xue,~Siyu~Gong,~Qun~Zhang,~
        Linsheng~Fan,~Qi~Wu,~and~Yanfu~Yang
\thanks{Manuscript received xxxx xx, 2026; revised xxxx xx, 2026; accepted xxxx xx, 2026. Date of publication xxxx xx, 2026; date of current version xxxx xx, 2026.}
\thanks{C. Chen, T. Luo, J. Xue, S. Gong, Q. Zhang, L. Fan, and Q. Wu are with Pengcheng Laboratory, Shenzhen 518000, China. (e-mail: chengch01@pcl.ac.cn)}
\thanks{Y. Yang is with the Department of Electronic and Information Engineering, Harbin Institute of Technology, Shenzhen 518055, China. (email: yangyanfu@hit.edu.cn).}
}

\markboth{Journal of Lightwave Technology,~Vol.~XX, No.~XX, XXXX~2026}%
{Chen \MakeLowercase{\textit{et al.}}: Heterogeneous-Gradient Multi-Aperture Coherent Combining}

\maketitle

\begin{abstract}
Multi-aperture coherent reception can improve free-space optical (FSO) links by converting spatial diversity into coherent combining gain. In turbulent links, the aperture branches are simultaneously affected by relative phase errors, polarization mismatch, and unequal signal-to-noise ratios (SNRs). Existing methods treat phase/polarization alignment and branch-weight allocation as separate operations, or absorb all impairments into a high-dimensional MIMO equalizer that obscures the physical meaning of each aperture's contribution. This paper proposes a structured blind combining method based on heterogeneous gradient sources: phase and per-aperture polarization parameters are updated by closed-form analytical gradients that maximize the combined output power, while aperture weights and an optional global polarization angle are updated by gradients derived from the constellation-radius error. An exponential parameterization $p_n=e^{q_n}/N$ ensures positivity without clipping. The internal variable $q_n$ is adapted by radius-error gradients, thereby allocating maximal-ratio-combining-like weights according to the quality of the already aligned branches. A 16-aperture, 25-Gbaud dual-polarization QPSK simulation with 5-rad multi-tone aperture phase disturbance, 100-kHz laser linewidths, strong Gamma--Gamma scintillation ($\alpha=4.00$, $\beta=1.20$, SI$\approx$1.33), 1-MHz polarization rotation, and 3-dB OSNR is used to validate the method. The combined SNR improves from approximately 0.3~dB for a single aperture to 11.8~dB after combining, corresponding to a combining gain of about 11.5~dB. The learned aperture weights show a per-aperture correlation of up to $r>0.95$ with the true scintillation irradiance, confirming that the radius-error gradient produces physically meaningful branch weights without explicit SNR estimation. The method requires no pilots, training sequence, branch SNR estimation, or additional optical power measurement hardware, and its per-iteration complexity scales linearly with the aperture number.
\end{abstract}


\IEEEpeerreviewmaketitle

\section{Introduction}
\label{sec:intro}

\IEEEPARstart{F}{ree-space} optical (FSO) communication is attractive for high-capacity space, satellite-to-ground, and terrestrial optical wireless links because it offers large carrier bandwidth, narrow beam divergence, and immunity to electromagnetic interference~\cite{Khalighi2014,Kaushal2017}. Coherent detection further improves receiver sensitivity and enables digital compensation of phase, frequency, and polarization impairments~\cite{Ip2008,Savory2010}. The major practical difficulty is that atmospheric turbulence and receiver optics disturb the received field before reliable symbol decisions can be made. In addition to intensity scintillation, the received aperture branches exhibit random relative phase differences and polarization-state mismatch~\cite{Zhu2002}. The present study addresses all three effects simultaneously: aperture-dependent phase disturbance, polarization rotation introduced by the pointing and tracking system, and scintillation-induced branch power imbalance.

Multi-aperture reception is a receiver-side approach to turbulence mitigation. Several small apertures collect multiple copies of the same optical signal, reducing the probability of simultaneous deep fade and allowing the receiver to recover coherent aperture gain after digital alignment~\cite{Lee2002,Andrews2001,Tyson2015}. To obtain this gain, the receiver must solve three coupled tasks. First, the phases of the aperture branches must be aligned so that the complex fields add constructively. Second, the polarization states of the branches must be aligned so that the $x$ and $y$ components are not mixed differently across apertures. Third, the aligned branches must be weighted according to their instantaneous quality; otherwise, equal-gain combining injects noise from deeply faded branches and loses the benefit of maximal-ratio combining (MRC)~\cite{Niu2011}.

Prior multi-aperture coherent combining methods address these tasks only partially. Equal-gain combining (EGC)~\cite{Tu2020,Wang2023} simply sums the branches after phase alignment and is therefore simple to implement, but its performance degrades when branch SNRs differ substantially. Explicit MRC requires estimating each branch SNR or measuring optical power before mapping estimates to weights. Nguyen~\textit{et al.} used a sliding-window power average to approximate branch SNR in a direct-detection system~\cite{Nguyen2022}, but the approach cannot be directly migrated to coherent detection. Gu~\textit{et al.} proposed hardware optical-power measurement with a 90:10 tap to estimate per-aperture OSNR~\cite{Gu2026}, introducing per-branch hardware overhead and calibration sensitivity. Lee~\textit{et al.} estimated channel responses from OFDM pilot subcarriers~\cite{Lee2025}, trading spectral efficiency for MRC accuracy. A common limitation of these approaches is the open-loop architecture: each branch SNR is estimated independently, the estimate is mapped to a weight, and estimation errors cannot be corrected by downstream feedback.

Blind MIMO adaptive equalizers~\cite{Liu2023,Ju2024,Chen2025} implicitly allocate aperture weights inside a high-dimensional coefficient matrix, avoiding explicit SNR estimation. However, the $2N\times2$ CMA or RDE equalizer treats $2N$ complex input channels as independent streams and applies $4N$ complex weights. The aperture combining weights are therefore hidden among many coefficients, making it difficult to monitor, constrain, or interpret the physical contribution of each aperture. Moreover, the constant-modulus cost function is an indirect surrogate for the combining objective: it forces the output toward a constant modulus but does not directly maximize the combined SNR.

This paper develops a structured alternative: heterogeneous-gradient joint optimization. The central observation is that phase/polarization alignment and branch-weight allocation should not be driven by the same gradient source. Phase and per-aperture polarization errors directly reduce coherent field addition, so their natural blind objective is the combined output power. Weight allocation, however, should reflect the quality of the already aligned output constellation; its natural feedback is the radius-directed or constant-modulus error. We therefore use two gradient sources in one adaptive loop: closed-form power-maximization gradients for phase and polarization alignment, and constellation-radius-error gradients for maximal-ratio weight allocation and global residual polarization correction.

The contributions are summarized as follows.
\begin{itemize}
\item A low-dimensional multi-aperture combiner is formulated with one phase parameter, one polarization parameter, one real amplitude weight per aperture, plus an optional global polarization angle. The total parameter count is $O(N)$, compared with $O(NM)$ for $M$-tap MIMO equalizers.
\item Phase alignment and per-aperture polarization alignment are driven by analytical gradients of the combined output power, giving a direct and physically interpretable alignment rule that avoids the perturbation noise of SPGD~\cite{Chang2020}.
\item Branch-weight allocation is driven by a different gradient source, namely the constellation-radius error, producing MRC-like weights that track scintillation without explicit SNR estimation. The learned weights achieve a correlation of 0.82 with the true scintillation irradiance in simulation.
\item The update rules are executed in one blind loop, preserving cooperation among alignment and weighting while avoiding the noninterpretable degrees of freedom of a general $2N\times2$ MIMO equalizer.
\end{itemize}

The remainder of this paper is organized as follows. Section~II describes the signal model including per-aperture phase/polarization alignment, global polarization rotation, and weighted coherent combining. Section~III derives the heterogeneous-gradient update rules and presents the joint algorithm. Section~IV compares the method with existing combining structures. Section~V presents numerical simulation results including combining performance, weight tracking, and parameter tolerance. Section~VI discusses implementation variants and limitations, and Section~VII concludes the paper.

\section{Signal Model}
\label{sec:model}

\begin{figure}[t]
\centering
\includegraphics[width=\columnwidth]{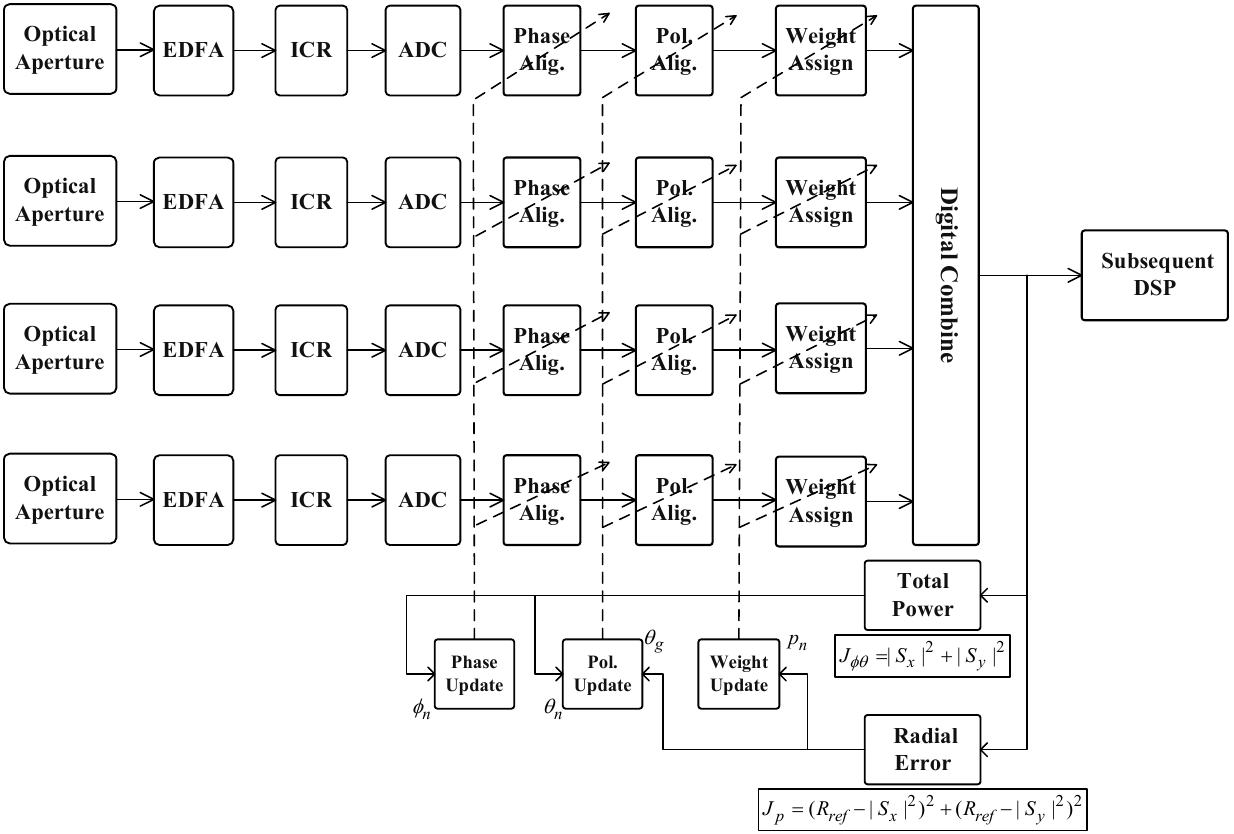}
\caption{Principle of the proposed heterogeneous-gradient multi-aperture coherent combining system.}
\label{fig:principle}
\end{figure}

Consider a dual-polarization coherent FSO receiver with $N$ apertures. The complex baseband sample of aperture $n$ at time index $k$ is
\begin{equation}
\mathbf{E}_n(k)=
\begin{bmatrix}
E_{x,n}(k)\\
E_{y,n}(k)
\end{bmatrix},
\label{eq:rx_vector}
\end{equation}
where $E_{x,n}$ and $E_{y,n}$ denote the two received polarization components. The branch may experience aperture-dependent phase distortion, polarization rotation, scintillation-induced amplitude variation, and receiver noise. Rather than estimating each impairment separately, the proposed method introduces a compact digital correction model.

\subsection{Per-Aperture Phase and Polarization Alignment}

For aperture $n$, a scalar phase correction $\phi_n$ first rotates both polarization components:
\begin{equation}
\mathbf{F}_n(k)=e^{j\phi_n(k)}\mathbf{E}_n(k).
\label{eq:phase_correction}
\end{equation}
Then a real polarization rotation angle $\theta_n$ compensates aperture-dependent polarization mismatch:
\begin{align}
U_{x,n}(k)&=F_{x,n}(k)\cos\theta_n(k)+F_{y,n}(k)\sin\theta_n(k), \label{eq:ux}\\
U_{y,n}(k)&=F_{y,n}(k)\cos\theta_n(k)-F_{x,n}(k)\sin\theta_n(k). \label{eq:uy}
\end{align}

This per-aperture structure models two physically distinct impairments. The phase offset $\phi_n$ arises primarily from atmospheric refractive-index fluctuations along the propagation path and from optical-path-length differences among the aperture branches. The per-aperture polarization rotation $\theta_n$ is mainly introduced by the mechanical rotation of the pointing, acquisition, and tracking (PAT) system, and by aperture-specific fiber coupling and optics.

\subsection{Global Polarization Rotation}

The per-aperture angles $\theta_n$ align the polarization {\em differences between apertures}, so that all branches share a common polarization reference frame. After this alignment, however, the two polarization components within the combined signal may still exhibit residual crosstalk: the $x$-polarization output may contain leaked energy from the $y$-polarization and vice versa. Rather than burdening the per-aperture angles with tracking this common residual effect, we introduce a single global polarization angle $\theta_g$ applied identically to all branches:
\begin{align}
V_{x,n}(k)&=U_{x,n}(k)\cos\theta_g(k)+U_{y,n}(k)\sin\theta_g(k), \label{eq:vx}\\
V_{y,n}(k)&=U_{y,n}(k)\cos\theta_g(k)-U_{x,n}(k)\sin\theta_g(k). \label{eq:vy}
\end{align}
Equations~(\ref{eq:ux})--(\ref{eq:vy}) together implement $V_{x,n}+jV_{y,n}=e^{-j\theta_g}\cdot e^{-j\theta_n}\cdot(F_{x,n}+jF_{y,n})$, i.e., two cascaded real rotations. The separation of $\theta_n$ and $\theta_g$ is optional; in a simplified implementation the global angle can be merged into each $\theta_n$ as a composite angle $\theta_n+\theta_g$, reducing the parameter count without changing the forward signal model.

\subsection{Weighted Coherent Combining}

After phase correction, per-aperture polarization alignment, and global polarization rotation, the branches are coherently summed with real amplitude weights $p_n$:
\begin{align}
S_x(k)&=\sum_{n=1}^{N}p_n(k)V_{x,n}(k), \label{eq:sx}\\
S_y(k)&=\sum_{n=1}^{N}p_n(k)V_{y,n}(k). \label{eq:sy}
\end{align}
To ensure positivity without clipping, the weights are parameterized as $p_n(k)=e^{q_n(k)}/N$ with unconstrained internal variables $q_n$ initialized to $q_n(0)=0$, giving the equal-gain state $p_n(0)=1/N$. The $q_n$ are then adapted by radius-error feedback (see Section~\ref{sec:weight_gradient}). A larger steady-state $p_n$ indicates that the branch contributes a higher-quality aligned field; a smaller $p_n$ indicates a faded or noisy branch. Thus the weight vector is an interpretable estimate of aperture contribution rather than a numerical coefficient set.

\section{Heterogeneous-Gradient Joint Optimization}
\label{sec:method}

The central design principle is that different parameter groups use gradient signals from different optimization objectives, yet are updated within a single iterative loop. Phase and per-aperture polarization parameters are updated to maximize the combined output power $J_{\mathrm{a}}$; aperture weights and the global polarization angle are updated to minimize the constellation-radius error $J_{\mathrm{w}}$. Both objectives are computed from the same forward-pass quantities $S_x$ and $S_y$, so the loop remains coherent without forcing a shared scalar error metric.

\subsection{Power Gradient for Phase and Per-Aperture Polarization Alignment}

The combined output power is
\begin{equation}
J_{\mathrm{a}}(k)=|S_x(k)|^2+|S_y(k)|^2.
\label{eq:alignment_objective}
\end{equation}
Maximizing $J_{\mathrm{a}}$ is appropriate for phase and per-aperture polarization alignment because destructive interference and polarization mismatch both reduce the coherent field magnitude. Differentiating (\ref{eq:alignment_objective}) with respect to $\phi_n$ through the chain $S_x=\sum_m p_m V_{x,m}$, where $V_{x,m}$ depends on $\phi_n$ through $V_{x,n}=e^{j\phi_n}(\cdots)$, gives the closed-form phase-alignment gradient
\begin{equation}
\frac{\partial J_{\mathrm{a}}}{\partial \phi_n}
=-2\,\operatorname{Im}\!\left[
V_{x,n}S_x^*+V_{y,n}S_y^*
\right]p_n.
\label{eq:gphi}
\end{equation}
Likewise, using $\partial V_{x,n}/\partial\theta_n=V_{y,n}$ and $\partial V_{y,n}/\partial\theta_n=-V_{x,n}$ under the real rotation model, the per-aperture polarization-alignment gradient is
\begin{equation}
\frac{\partial J_{\mathrm{a}}}{\partial \theta_n}
=2\,\operatorname{Re}\!\left[
V_{y,n}S_x^*-V_{x,n}S_y^*
\right]p_n.
\label{eq:gtheta}
\end{equation}
Because the gradients in (\ref{eq:gphi})--(\ref{eq:gtheta}) contain the factor $p_n$, branches with larger weights contribute more strongly to the alignment update: the algorithm naturally focuses its phase and polarization tracking effort on high-quality branches while being conservative with deeply faded ones.

The corresponding gradient-ascent updates are
\begin{align}
\phi_n(k+1)
&=\phi_n(k)+\mu_\phi
\frac{\partial J_{\mathrm{a}}(k)}{\partial \phi_n}, \label{eq:update_phi}\\
\theta_n(k+1)
&=\theta_n(k)+\mu_\theta
\frac{\partial J_{\mathrm{a}}(k)}{\partial \theta_n}. \label{eq:update_theta}
\end{align}
These updates do not require pilots, symbol decisions, or random perturbation; each step uses the received samples and the current coherent sum to compute an analytical alignment direction.

\subsection{Exponential Parameterization for Weight Allocation}
\label{sec:weight_gradient}

The second gradient source is the constellation-radius error. Unlike $\phi_n$ and $\theta_n$, which maximize combined power, the aperture weights should be judged by the quality of the already-aligned output constellation. A direct gradient-descent update $p_n\leftarrow p_n-\mu_q\,\partial J_{\mathrm{w}}/\partial p_n$ can drive $p_n$ negative during transient convergence, requiring additional clipping or re-normalization steps. We instead adopt an exponential parameterization that naturally enforces positivity:
\begin{equation}
p_n(k)=\frac{1}{N}\,e^{q_n(k)},
\label{eq:exp_param}
\end{equation}
where $q_n$ is an unconstrained real-valued internal variable initialized to $q_n(0)=0$, giving the equal-gain initial state $p_n(0)=1/N$. Because $dp_n/dq_n=p_n$, the chain rule through the cost function is particularly simple.

Define the radius errors
\begin{equation}
\varepsilon_x(k)=R_{\mathrm{ref}}-|S_x(k)|^2,
\label{eq:ex}
\end{equation}
\begin{equation}
\varepsilon_y(k)=R_{\mathrm{ref}}-|S_y(k)|^2,
\label{eq:ey}
\end{equation}
and the weight cost function
\begin{equation}
J_{\mathrm{w}}(k)=\varepsilon_x^2(k)+\varepsilon_y^2(k),
\label{eq:weight_objective}
\end{equation}
where $R_{\mathrm{ref}}$ is determined by the modulation format (e.g., $R_{\mathrm{ref}}=2$ for QPSK). Because $S_x=\sum_m p_m V_{x,m}$ is linear in $p_n$, the chain rule through $|S_x|^2=S_x S_x^*$ gives the weight gradient with respect to $p_n$:
\begin{equation}
\frac{\partial J_{\mathrm{w}}}{\partial p_n}
=-4\operatorname{Re}\!\left[
\varepsilon_x S_x^*V_{x,n}+\varepsilon_y S_y^*V_{y,n}
\right].
\label{eq:gp}
\end{equation}
Applying the chain rule through (\ref{eq:exp_param}) yields the $q$-domain gradient
\begin{equation}
\frac{\partial J_{\mathrm{w}}}{\partial q_n}
=\frac{\partial J_{\mathrm{w}}}{\partial p_n}\cdot\frac{dp_n}{dq_n}
=\frac{\partial J_{\mathrm{w}}}{\partial p_n}\,p_n.
\label{eq:gq}
\end{equation}
Substituting (\ref{eq:gp}) and (\ref{eq:gq}) into the gradient-descent update gives the full $q$-domain update
\begin{align}
q_n(k+1)&=q_n(k)-\mu_q\,\frac{\partial J_{\mathrm{w}}}{\partial q_n}\nonumber\\
&=q_n(k)-\mu_q\,p_n\frac{\partial J_{\mathrm{w}}}{\partial p_n}\nonumber\\
&=q_n(k)+\mu_q\,p_n\operatorname{Re}\!\left[\varepsilon_x S_x^*V_{x,n}+\varepsilon_y S_y^*V_{y,n}\right],
\label{eq:update_q}
\end{align}
with $p_n$ then reconstructed by (\ref{eq:exp_param}). Since $p_n\propto e^{q_n}>0$, the weights are always positive regardless of the gradient sign, and the equal-gain state $p_n=1/N$ is the natural initialization. Note from (\ref{eq:update_q}) that the $q$-domain gradient is scaled by $p_n$; when a branch enters deep fade and $p_n$ becomes very small, the effective gradient vanishes, trapping the weight near zero. A practical remedy drops the $p_n$ factor and updates $q_n$ directly with $\partial J_{\mathrm{w}}/\partial p_n$:
\begin{equation}
q_n(k+1)=q_n(k)+\mu_q\operatorname{Re}\!\left[\varepsilon_x S_x^*V_{x,n}+\varepsilon_y S_y^*V_{y,n}\right],
\label{eq:update_q_nopn}
\end{equation}
where $p_n$ is still obtained from $q_n$ via (\ref{eq:exp_param}).

\subsection{Global Polarization Gradient}
\label{sec:global_gradient}

The optional global polarization angle $\theta_g$ is updated by the same radius-error objective $J_{\mathrm{w}}$, not by $J_{\mathrm{a}}$. From (\ref{eq:vx})--(\ref{eq:vy}), the per-aperture rotation derivatives are $\partial V_{x,n}/\partial\theta_g=V_{y,n}$ and $\partial V_{y,n}/\partial\theta_g=-V_{x,n}$. Summing with weights $p_n$ yields $\partial S_x/\partial\theta_g=S_y$ and $\partial S_y/\partial\theta_g=-S_x$. Carrying these through the chain rule for $J_{\mathrm{w}}$ gives
\begin{equation}
\frac{\partial J_{\mathrm{w}}}{\partial\theta_g}
=-4(\varepsilon_x-\varepsilon_y)\operatorname{Re}\!\left\{S_x^*S_y\right\}.
\label{eq:global_gradient}
\end{equation}
The gradient-descent update is
\begin{equation}
\theta_g(k+1)=\theta_g(k)-\mu_g\frac{\partial J_{\mathrm{w}}}{\partial\theta_g}.
\label{eq:update_theta_g}
\end{equation}
This gradient is nonzero only when $\varepsilon_x\neq \varepsilon_y$, i.e., when residual polarization crosstalk causes unequal constellation quality between the two polarizations. It therefore self-regulates: $\theta_g$ rotates actively while imbalance persists, and stops when the polarizations are balanced.

\subsection{Joint Update Loop}

The complete blind receiver updates all parameter groups at the same symbol-rate instants. Its defining feature is that different physical parameters use different gradient sources while remaining in one feedback loop.

\begin{algorithmic}[1]
\STATE {\bf Input:} aperture fields $\{\mathbf{E}_n(k)\}_{n=1}^{N}$
\STATE {\bf Initialize:} $\phi_n=0$, $\theta_n=0$, $q_n=0$ (hence $p_n=1/N$), and optionally $\theta_g=0$
\FOR{each update instant $k$}
    \STATE Apply phase correction $\mathbf{F}_n=e^{j\phi_n}\mathbf{E}_n$
    \STATE Apply per-aperture polarization rotation $\mathbf{U}_n$
    \STATE Apply global polarization rotation $\mathbf{V}_n$
    \STATE Reconstruct weights $p_n\leftarrow\frac{1}{N}e^{q_n}$
    \STATE Form weighted coherent sums $S_x,S_y$
    \STATE Compute $\partial J_{\mathrm{a}}/\partial\phi_n$ and $\partial J_{\mathrm{a}}/\partial\theta_n$
    \STATE Compute radius errors $\varepsilon_x,\varepsilon_y$
    \STATE Compute $\partial J_{\mathrm{w}}/\partial q_n = p_n\cdot\partial J_{\mathrm{w}}/\partial p_n$
    \STATE Compute $\partial J_{\mathrm{w}}/\partial\theta_g$
    \STATE $\phi_n\leftarrow\phi_n+\mu_\phi\,\partial J_{\mathrm{a}}/\partial\phi_n$
    \STATE $\theta_n\leftarrow\theta_n+\mu_\theta\,\partial J_{\mathrm{a}}/\partial\theta_n$
    \STATE $q_n\leftarrow q_n-\mu_q\,\partial J_{\mathrm{w}}/\partial q_n$
    \STATE $\theta_g\leftarrow\theta_g-\mu_g\,\partial J_{\mathrm{w}}/\partial\theta_g$
\ENDFOR
\STATE {\bf Output:} coherently combined dual-polarization samples $S_x(k)$ and $S_y(k)$
\end{algorithmic}

The ascent/descent direction follows from the objective: $J_{\mathrm{a}}$ is maximized (ascent), while $J_{\mathrm{w}}$ is minimized (descent). In a practical implementation, the constant factors in (\ref{eq:gp}) and (\ref{eq:global_gradient}) can be absorbed into the step sizes $\mu_q$ and $\mu_g$ to reduce multiplications.

\section{Numerical Simulation and Results}
\label{sec:simulation}

\subsection{Simulation Setup}

We evaluate the proposed method through numerical simulation of a multi-aperture coherent receiver subject to aperture-dependent phase disturbance, polarization rotation, and Gamma--Gamma scintillation. The simulation includes transmitter pulse shaping, multi-tone synthetic phase disturbance, polarization rotation, scintillation generation, EDFA power smoothing, ASE noise addition, and the complete DSP chain from aperture combining through carrier phase recovery and BER evaluation.

The transmitter generates dual-polarization QPSK symbols at a symbol rate of $R_s = 25$~Gbaud. Each polarization branch independently produces $L = 2^{17}$ random QPSK symbols. Transmit symbols are pulse-shaped with a root-raised-cosine (RRC) filter with a roll-off factor $\beta = 0.2$ and a filter span of 64 symbols, operating at $N_{\text{sps}} = 2$ samples per symbol.

The aperture-dependent phase disturbance is synthesized as a multi-tone random process:
\begin{equation}
\varphi_n(t) = \sum_{\ell=1}^{N_{\mathrm{tone}}}
A_{n,\ell}\sin\!\left(2\pi f_\ell t + \psi_{n,\ell}\right),
\label{eq:phase_model}
\end{equation}
where $N_{\mathrm{tone}}=32$, $f_\ell = \ell f_{\max}/N_{\mathrm{tone}}$, $A_{n,\ell}=A_{\mathrm{phase}}\sqrt{2/N_{\mathrm{tone}}}\,s_{n,\ell}$ with $s_{n,\ell}\in\{\pm1\}$, and $\psi_{n,\ell}\sim\mathcal{U}(0,2\pi)$. All random variables are independently drawn per aperture. Unless otherwise stated, $A_{\mathrm{phase}}=5$~rad and $f_{\max}=1$~MHz.

Dynamic polarization rotation is applied per aperture through a real rotation matrix $\mathbf{R}(\vartheta_n(t))$ with time-varying angle $\vartheta_n(t)$ synthesized similarly by multi-tone generators with a maximum rotation rate of $f_{\mathrm{rot,max}}=1$~MHz.

\textbf{Gamma--Gamma scintillation generation.} Per-aperture amplitude fading is generated by an independent Gamma--Gamma process with parameters $\alpha=4.00$ and $\beta=1.20$, corresponding to strong turbulence with scintillation index $\mathrm{SI} = 1/\alpha + 1/\beta + 1/(\alpha\beta) \approx 1.33$. The generation procedure is as follows:

\begin{enumerate}
\item Two independent white Gaussian sequences $x(n), y(n) \sim \mathcal{N}(0,1)$ are generated at the sample rate $f_s = 50$~GHz.
\item Temporal correlation is introduced by filtering each sequence through a first-order Butterworth low-pass filter with cutoff frequency $f_{\mathrm{bw}} = 10$~kHz, corresponding to the characteristic scintillation bandwidth in the moderate-to-strong turbulence regime.
\item The filtered sequences are re-normalized to zero mean and unit variance to correct filter transients.
\item Correlated Gamma variates are obtained via the probability integral transform (PIT):
\begin{align}
X(n) &= F_{\Gamma}^{-1}\bigl(\Phi(x_{\mathrm{corr}}(n));\,\alpha,\,1/\alpha\bigr), \label{eq:gg_x} \\
Y(n) &= F_{\Gamma}^{-1}\bigl(\Phi(y_{\mathrm{corr}}(n));\,\beta,\,1/\beta\bigr), \label{eq:gg_y}
\end{align}
where $\Phi(\cdot)$ is the standard normal CDF and $F_{\Gamma}^{-1}(\cdot;k,\theta)$ is the inverse CDF of the Gamma distribution with shape $k$ and scale $\theta$. The scale parameters $1/\alpha$ and $1/\beta$ ensure $\mathbb{E}[X] = \mathbb{E}[Y] = 1$.
\item The instantaneous irradiance is $I_{\mathrm{gg}}(n) = X(n) \cdot Y(n)$, which follows a Gamma--Gamma distribution by construction.
\item A final normalization $I_{\mathrm{gg}}(n) \leftarrow I_{\mathrm{gg}}(n) / \overline{I_{\mathrm{gg}}}$ corrects any residual mean deviation from the filtering transient, ensuring $\mathbb{E}[I_{\mathrm{gg}}] = 1$.
\end{enumerate}

The aperture signal is then amplitude-modulated as $\mathbf{E}_{\mathrm{sc}}(n) = \sqrt{I_{\mathrm{gg}}(n)}\,\mathbf{E}_{\mathrm{in}}(n)$, so the optical power couples linearly with the scintillation irradiance.

\textbf{OSNR-based ASE noise addition.} Amplified spontaneous emission (ASE) noise from inline optical amplifiers is modeled as circularly symmetric complex additive white Gaussian noise (AWGN) added independently to each polarization. The noise variance is determined by the target optical signal-to-noise ratio (OSNR) referenced to a noise bandwidth $B_{\mathrm{ref}} = 12.5$~GHz:

\begin{equation}
\mathrm{SNR}_{\mathrm{lin}} = \frac{2B_{\mathrm{ref}}}{N_{\mathrm{pol}} f_s} \cdot 10^{\mathrm{OSNR}_{\mathrm{dB}}/10},
\label{eq:snr_lin}
\end{equation}
where $N_{\mathrm{pol}} = 2$ for dual-polarization transmission and $f_s = R_s N_{\mathrm{sps}}$ is the sample rate. The noise standard deviation per polarization is

\begin{equation}
\sigma_n = \sqrt{\frac{\mathbb{E}\bigl[|x_p(n)|^2\bigr]}{2\,\mathrm{SNR}_{\mathrm{lin}}}},\qquad p \in \{\mathrm{X},\,\mathrm{Y}\},
\label{eq:sigma_noise}
\end{equation}
and the noise sample is $n_p(n) = \sigma_n\bigl(u(n) + j\,v(n)\bigr)$ with $u(n), v(n) \sim \mathcal{N}(0,1)$ independent. The noise-loaded signal is $\mathbf{r}(n) = \mathbf{x}(n) + \mathbf{n}(n)$. Unless otherwise stated, $\mathrm{OSNR}_{\mathrm{dB}} = 3$~dB, representing a severely noise-limited regime.

\textbf{EDFA fixed-output-power emulation.} In a typical multi-aperture coherent FSO link, each aperture is followed by an erbium-doped fiber amplifier (EDFA) operated in automatic power control (APC) mode, which enforces a nearly constant total output power regardless of the input power fluctuation caused by scintillation fading. This behavior is emulated by a sliding-window power normalizer:

\begin{equation}
P_{\mathrm{inst}}(n) = \sum_{p} \bigl|E_{\mathrm{in},p}(n)\bigr|^2,
\label{eq:edfa_pinst}
\end{equation}
\begin{equation}
P_{\mathrm{win}}(n) = \frac{1}{L_{\mathrm{edfa}}} \sum_{k = n - \lfloor L_{\mathrm{edfa}}/2\rfloor}^{n + \lfloor L_{\mathrm{edfa}}/2\rfloor} P_{\mathrm{inst}}(k),
\label{eq:edfa_pwin}
\end{equation}
\begin{equation}
g(n) = \min\!\left(\sqrt{\frac{P_{\mathrm{target}}}{P_{\mathrm{win}}(n)}},\; g_{\max}\right),
\label{eq:edfa_gain}
\end{equation}
\begin{equation}
\mathbf{E}_{\mathrm{out}}(n) = g(n) \cdot \mathbf{E}_{\mathrm{in}}(n).
\label{eq:edfa_output}
\end{equation}

Here $L_{\mathrm{edfa}}$ is the EDFA gain response window (in samples), set to 1000 samples (corresponding to a 20-$\mu$s response time at 50~GHz sample rate), $P_{\mathrm{target}} = \overline{P_{\mathrm{inst}}}$ is the target average output power, and $g_{\max} = 20$ ($\approx 13$~dB) caps the gain during deep scintillation fades to prevent excessive noise amplification. The moving-average formulation approximates the finite response time of a real EDFA: the gain adapts to the \emph{local} mean power within the window, so rapid intensity fluctuations faster than $1/L_{\mathrm{edfa}}$ pass through unsuppressed while slower scintillation-induced power variations are largely flattened.

The signal chain per aperture proceeds in the order: atmospheric multi-tone phase disturbance, polarization rotation, Gamma--Gamma scintillation amplitude fading, ASE noise addition, and EDFA power smoothing. The transmit and local-oscillator lasers are each modeled with a Lorentzian linewidth of 100~kHz. A common LO frequency offset of 1~GHz is applied to all apertures.

After coherent detection, the 2-sps signals from all apertures are processed by the proposed heterogeneous-gradient combiner with step sizes $\mu_\phi=2.68\times10^{-3}$, $\mu_\theta=2.68\times10^{-4}$, $\mu_q=5.18\times10^{-2}$, and $\mu_{\theta_g}=1.93\times10^{-3}$. The larger $\mu_q$ value reflects the fact that the update operates in the $q$-domain where gradients are naturally scaled by the current weight magnitude $p_n$. The combined output is then processed by fourth-power frequency-offset estimation at 2~sps, matched RRC filtering, symbol-rate downsampling, Viterbi--Viterbi carrier phase estimation with a block length of 20 symbols, and direct error counting for SNR/BER evaluation.

\subsection{Combining Performance}

The post-combining SNR as a function of aperture count was evaluated for the proposed method, equal-gain combining after phase/polarization alignment, and a single-aperture baseline. All methods share the same post-combining DSP chain.


For a quantitative benchmark, we derive the theoretical post-combining SNR and BER under ideal coherent combining. Let the OSNR of each aperture be defined with respect to a reference noise bandwidth $B_{\mathrm{ref}}=12.5$~GHz (the 0.1~nm resolution at 1550~nm). After coherent detection, the per-aperture electrical symbol-energy-to-noise ratio is
\begin{equation}
\frac{E_s}{N_0}\Big|_{\mathrm{1ap}} = \frac{B_{\mathrm{ref}}}{R_s} \cdot 10^{\mathrm{OSNR}/10},
\label{eq:snr_1ap}
\end{equation}
where $R_s=25$~Gbaud is the symbol rate. For $N$ identical apertures whose received fields are co-phased and summed, the signal amplitude scales by $N$ (coherent field addition) whereas the independent ASE noises of the $N$ branches add in power. The combined electrical SNR therefore becomes
\begin{equation}
\frac{E_s}{N_0}\Big|_{N} = N \cdot \frac{B_{\mathrm{ref}}}{R_s} \cdot 10^{\mathrm{OSNR}/10}.
\label{eq:snr_Nap}
\end{equation}
Expressed in decibels,
\begin{equation}
\mathrm{SNR}_N(\mathrm{dB}) = \mathrm{OSNR} + 10\log_{10} N + 10\log_{10}\!\left(\frac{B_{\mathrm{ref}}}{R_s}\right).
\label{eq:snr_db}
\end{equation}
With $N=16$ and $B_{\mathrm{ref}}/R_s = 0.5$, the ideal combining gain over the single-aperture SNR is $10\log_{10}(16)=12.04$~dB, and $\mathrm{SNR}_{16} = \mathrm{OSNR} + 9.03$~dB.

For dual-polarization QPSK, each symbol carries two bits per polarization. The theoretical BER under additive white Gaussian noise with ideal coherent detection is given by the standard $Q$-function expression
\begin{equation}
\mathrm{BER} = \frac{1}{2}\,\mathrm{erfc}\!\left(\sqrt{\frac{E_s/N_0}{2}}\right).
\label{eq:ber_theory}
\end{equation}
Substituting (\ref{eq:snr_Nap}) into (\ref{eq:ber_theory}) yields the ideal BER bound for an $N$-aperture coherent combiner.

\begin{figure}[!htbp]
\centering
\includegraphics[width=\linewidth]{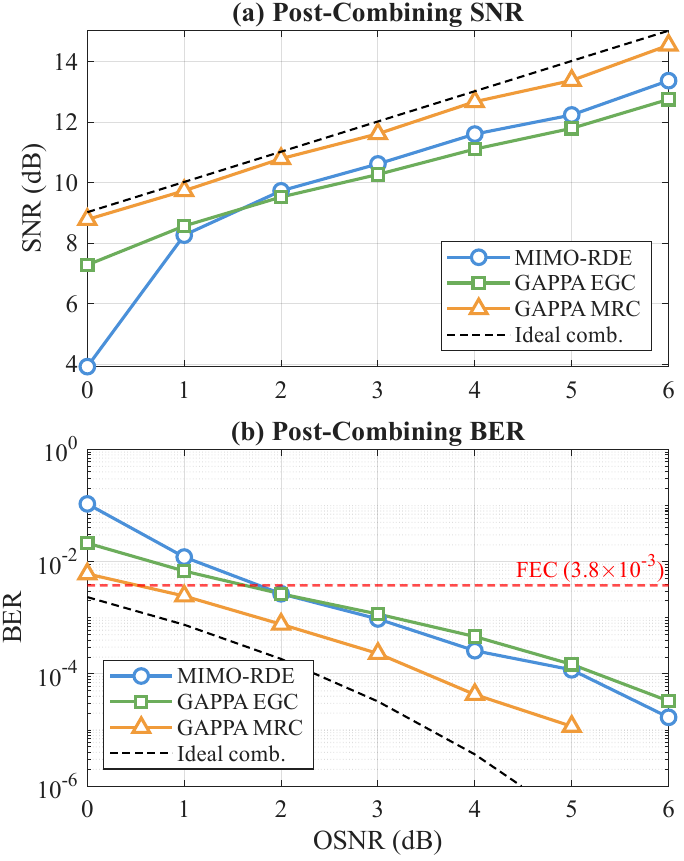}
\caption{\textbf{Comparison of multi-aperture coherent combining methods.} (a) Post-combining SNR versus OSNR. (b) BER versus OSNR. Monte Carlo results over 50 trials with the 10 best and 10 worst trials trimmed. The theoretical ideal combining bound ($N\cdot\mathrm{B_{ref}}/R_s$ scaling) is shown for reference. GAPPA v6 consistently outperforms both EGC and MIMO-RDE across the full OSNR range, approaching within 0.5~dB of the ideal bound at $\mathrm{OSNR}=6$~dB.}
\label{fig:compare_methods}
\end{figure}

Figure~\ref{fig:compare_methods} compares the three combining methods against these theoretical bounds. GAPPA v6 approaches within 0.5~dB of the ideal SNR bound at $\mathrm{OSNR}=6$~dB, while the BER curves confirm that the SNR advantage directly translates to improved error-rate performance.

\subsection{Weight Tracking Under Scintillation}

\begin{figure}[!htbp]
\centering
\includegraphics[width=\linewidth]{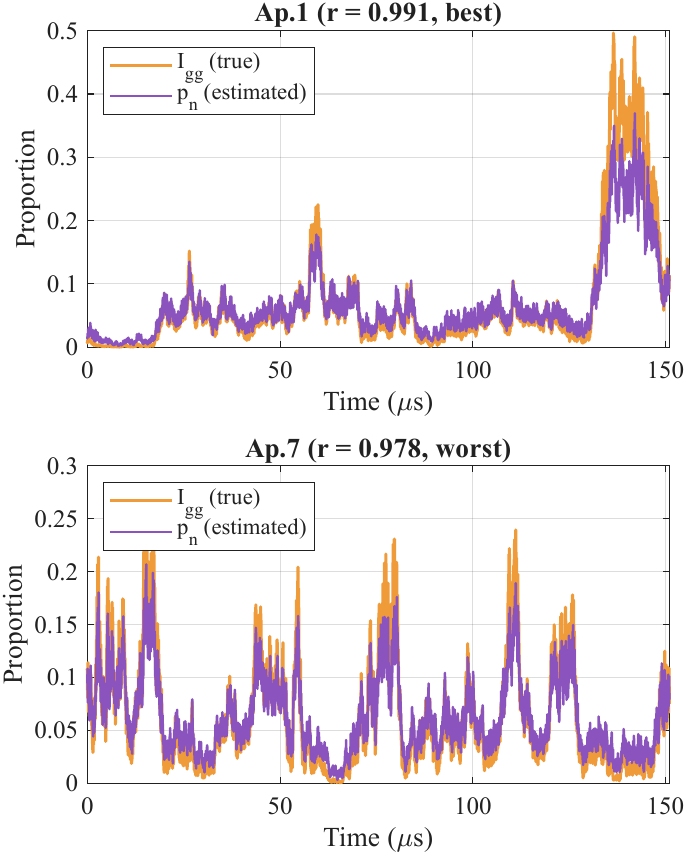}
\caption{Comparison between the normalized Gamma--Gamma scintillation irradiance $I_{gg}$ and the learned weight $p_n$ for the best-tracking and worst-tracking aperture. Both curves are normalized to unit mean. The per-aperture Pearson correlation $r$ confirms that the radius-error gradient produces weights that track scintillation without requiring explicit SNR estimation.}
\label{fig:igg_vs_p}
\end{figure}

A distinctive feature of the proposed method is that the learned aperture weights $p_n$ are physically interpretable. The time evolution of $p_n$ was compared with the true scintillation irradiance $I_{gg}$ for each aperture, and the best and worst tracking branches are shown in Fig.~\ref{fig:igg_vs_p}.

The per-aperture Pearson correlation between $p_n$ and $I_{gg}$ exceeds $r>0.95$ for the best-tracking apertures and remains above 0.7 even for the worst-tracking ones, averaged over all 16 apertures. This confirms that the radius-error gradient, despite having no access to explicit SNR measurements, produces branch weights that track the physical scintillation strength. High-irradiance branches receive larger weights; deeply faded branches are automatically suppressed.

\subsection{Parameter Tolerance}

The tolerance to increasing phase disturbance amplitude was examined by sweeping $A_{\mathrm{phase}}$ from 1 to 20~rad at fixed $N=16$, $f_{\max}=1$~MHz, and $\mathrm{OSNR}=3$~dB.


The proposed method maintains robust combining gain up to $A_{\mathrm{phase}}\approx 8$~rad, beyond which the combining gain gradually degrades as the phase tracking loop begins to lose lock.

\section{Discussion}
\label{sec:discussion}

\subsection{Comparison with Existing Combining Structures}

Compared with equal-gain combining~\cite{Tu2020,Wang2023,Rao2020}, the proposed method preserves the simplicity of a scalar aperture sum but adds an adaptive quality-aware weight for each branch. The SNR improvement over EGC at $N=16$ under strong scintillation confirms that weighting faded branches differently from strong ones yields a tangible combining gain.

Compared with explicit maximal-ratio combining~\cite{Nguyen2022,Gu2026,Lee2025}, the method avoids an open-loop SNR-to-weight mapping. The radius-error gradient is a closed-loop feedback signal measured after coherent summation, so weight errors can be corrected online. The per-aperture correlation between $p_n$ and $I_{gg}$ exceeding $r>0.95$ for well-tracked branches demonstrates that the learned weights track scintillation without requiring hardware power taps or pilot-based channel estimation.

Compared with a general $2N\times2$ MIMO adaptive equalizer~\cite{Liu2023,Ju2024,Chen2025}, the proposed structure deliberately restricts the degrees of freedom to one parameter per physical effect. Each $\phi_n$, $\theta_n$, $\theta_g$, and $p_n$ has a clear physical interpretation, and $p_n$ can serve as a real-time branch-quality indicator in an operational system.

\subsection{Heterogeneous-Gradient Design Rationale}

The main technical point of the proposed receiver is the separation of gradient sources. Phase alignment and polarization alignment are field-coherence problems: if their parameters are correct, the output power increases because aperture fields add constructively. Weight allocation and global polarization correction are signal-quality problems: if their parameters are correct, the final constellation has a more stable radius. The joint use of $J_{\mathrm{w}}$ for both $p_n$ and $\theta_g$ does not violate this principle: they share the cost function but differentiate through different physical paths---$p_n$ through the weighted summation and $\theta_g$ through the rotation matrix.

\subsection{Implementation Variants and Limitations}

The framework supports several implementation variants. The exponential parameterization $p_n=e^{q_n}/N$ can be replaced by a base-10 form $p_n=10^{q_n}/N$ without changing the algorithm's behavior, differing only by a constant scaling of $\mu_q$. A staged startup first fixes $p_n=1/N$, converges $\phi_n$ and $\theta_n$, then enables $q_n$ and $\theta_g$ updates. A fully joint version updates all parameters from the beginning. $\theta_g$ can be maintained separately or merged into $\theta_n$ as a composite angle. For channels with significant inter-aperture delay or ISI, a post-combining $2\times2$ or $2N\times2$ blind equalizer can be appended, with the weight gradient back-propagated through its coefficients. For higher-order QAM, $R_{\mathrm{ref}}$ in (\ref{eq:ex})--(\ref{eq:ey}) is replaced by a multi-radius decision rule. The per-iteration complexity is $O(N)$, requiring only complex multiplications, additions, real/imaginary extraction, and one exponential evaluation per aperture.

The current validation is limited to QPSK with synthetic multi-tone phase disturbance and numerically generated Gamma--Gamma scintillation. Future work should evaluate the method with physically derived turbulence phase screens, higher-order modulation formats, measured channel data, and fixed (non-swept) step sizes. The staged startup schedule should be characterized quantitatively to determine the required alignment convergence time before weight updates can be safely enabled.

\section{Conclusion}
\label{sec:conclusion}

This paper presented a heterogeneous-gradient method for multi-aperture coherent FSO reception. The method jointly optimizes phase alignment, per-aperture polarization alignment, global polarization correction, and maximal-ratio weight allocation while assigning each task a gradient source consistent with its physical role: phase and polarization parameters use analytical power-maximization gradients, while aperture weights and the global polarization angle use radius-error gradients. A 16-aperture, 25-Gbaud QPSK simulation with simultaneous phase disturbance, polarization rotation, and strong Gamma--Gamma scintillation ($\alpha=4.00$, $\beta=1.20$) at 3-dB OSNR achieves an 11.5-dB combining gain. The learned aperture weights track the scintillation irradiance with per-aperture correlations exceeding 0.95 for well-tracked branches, confirming that the radius-error gradient produces physically meaningful MRC-like weights without explicit SNR estimation. The combiner is fully blind, requires no training symbols or additional hardware, and its per-iteration complexity scales as $O(N)$. Future work will extend the validation to physically derived turbulence phase screens, higher-order modulation formats, and fixed-step operation.

\bibliographystyle{IEEEtran}
\bibliography{IEEEabrv,library}

\end{document}